\documentclass[leqno,authoryear,1p]{article}
\usepackage[utf8]{inputenc}
\usepackage{xcolor}
\usepackage{times}
\usepackage{enumitem}
\usepackage{bbold}
\usepackage{dsfont}
\usepackage{comment}
\usepackage{mathrsfs}
\usepackage{slashed}
\usepackage{soul}
\usepackage{amssymb,amsmath}
\usepackage{natbib}
\title{Is the world made of loops?}
\author{Alexander Afriat}
\begin{document}
\maketitle
\begin{abstract}
\noindent I see no good reason to prefer (any version I know of) the `holonomy interpretation' to the `potential interpretation' of the Aharonov-Bohm effect. Everyone agrees that the inverse image $[A]=[A+d\lambda]_{\lambda}=d^{-1}F$ of the electromagnetic field $F$ is a class, full of individuals; and that the circulation $\small{\textsf{C}}$ of the electromagnetic potential $A$ around a loop $\sigma_0$ encircling the solenoid is common to the whole class $[A]$, and to the homotopy class or \emph{hoop} $[\sigma_0]$. If picking individuals out of classes is the problem, picking an individual potential out of $[A]$ should be no worse than picking an individual loop out of $[\sigma_0]$. The individuals of $[A]$ can moreover be transcended---punctually, without integration around loops---by an appropriate version of the electromagnetic connection.

\end{abstract}

\section{Introduction}

Thales, one gathers, had nothing but \emph{water}; then came atoms, fire, air, earth, effluvia, fields, energy, waves and other complications. The history of ontological speculation (to say nothing of my garden, \S\ref{garden}) has now been enriched by \emph{loops}---and perhaps other boundaries too (\S\ref{electrostatics}).

The Aharonov-Bohm effect\footnote{\citet{EhrenbergSiday}, \citet{AharonovBohm}; see also \citet{Franz39, Franz40, Franz65}, \citet{OlariuPopescu}, \citet{Hiley}.} (\S\ref{AharonovBohm}) involves a relationship between variations in the current through a solenoid and changes in the interference pattern on a screen. The relationship is puzzling, but one can try to make sense of it in various ways. Of the available elements (electromagnetic field $F$ in the solenoid, loops enclosing the solenoid, wavefunction $\psi$, electromagnetic potential $A$, its circulation $\small{\textsf{C}}$ around the loops, interference pattern $\mathfrak{P}$, topological features), some can be chosen, others left out---with philosophical tradeoffs. Accounts emphasising the relationship between $\psi$ or $\mathfrak{P}$ and $F$ (rather than $A$) are disturbingly nonlocal.\footnote{\citet{AharonovBohm} p.~490: ``we might try to formulate a nonlocal theory in which, for example, the electron could interact with a field that was a finite distance away. Then there would be no trouble in interpreting these results, but, as is well known, there are severe difficulties in the way of doing this.'' See also \citet{Healey97}.} The problem can be overcome\footnote{\citet{AharonovBohm} pp.~490-1: ``we may retain the present local theory and [\,\dots] try to give a further new interpretation to the potentials. In other words, we are led to regard $A_{\mu}\textrm{(}x\textrm{)}$ as a physical variable.'' See also \citet{Feynman} \S15-5.} by recognising the existence of something like $A$ or $[A]$ or the electromagnetic connection; but then there's the (nontrivial) kernel of $d:A\mapsto F=dA$, which corresponds to the freedom
\begin{equation}\label{freedom}
[A]=[A+d\lambda]_{\lambda}=d^{-1}F.
\end{equation}
In the literature one finds enumerations of three or even four interpretations, which we can call:
\begin{enumerate}[label=(I\arabic*)]
\item \emph{electromagnetic field interpretation}
\item \emph{electromagnetic potential interpretation}
\item \emph{holonomy\footnote{The first two letters of ``anholonomy'' don't seem semantically irrelevant, quite on the contrary; the ``a'' looks very much like a transliterated alpha privative, which far from doing nothing at all would turn ``holonomy'' into its opposite, ``$\neg$ holonomy.'' One might imagine that the removal of the same two letters would restore the meaning of ``holonomy.'' Not at all; by a prodigy of language and logic we have
$$
\textrm{holonomy}\equiv\neg\textrm{ holonomy.}
$$
``Holonomy'' is often preferred to its anto-synonym (\emph{énantiosème} in French) ``anholonomy.'' Whether ``holonomy'' means ``anholonomy'' or the opposite is settled by context. I would only add to the confusion if I called the interpretation based on anholonomies---departures from holonomy---the ``anholonomy interpretation.''} interpretation}
\item \emph{topology interpretation}.
\end{enumerate}
The topology interpretation (I4), dealt with in \citet{THS}, will not be considered here. One can think of an annulus\footnote{One may prefer to think in two or three dimensions; not knowing what to call a three-dimensional annulus (``perforated cylinder''? ``solid annulus''?) I'll just say ``annulus.''} $\alpha$ between the solenoid and the surrounding wavefunction;\footnote{More details about the Aharonov-Bohm effect in \S\ref{AharonovBohm}. But it may already be helpful to imagine not one but two concentric cylindrical shields: one keeping $F$ \emph{in} inside another keeping $\psi$ \emph{out}. To bring out the nonlocal effect the distance, the annulus between the two shields can be made arbitrarily large.} the first three interpretations can be roughly distinguished by what they put in $\alpha$:
\begin{enumerate}[label=(I\arabic*)]
\item nothing at all
\item electromagnetic potential
\item loops.
\end{enumerate}
Interpretations (I1) and (I3) are so embarrassed by the \emph{embarras de richesses} (\ref{freedom}) that they altogether renounce the riches. In (I1) an empty annulus is simply accepted, whatever the consequences (again, a non-local influence between the electromagnetic field in the solenoid and the wavefunction beyond the annulus); in (I3) the void is filled with other embarrassing riches which are confidently passed off as altogether unembarrassing. In any case it is better to \emph{over}fill an awkward void, somehow or other, than to leave it empty; so (I2) and (I3) are both preferable to (I1). This paper is about (I3), as opposed to (I2) in particular. Again, I see no reason to prefer (I3) to (I2); the reason (\emph{embarras de richesses}) given for looking beyond (I2) to (I3) pertains to both, and is if anything more problematic in (I3) than in (I2).

I often have trouble making up my mind, and can see that \emph{having options} isn't always a good thing. But one mustn't exaggerate: having options (as with $[A]$) can't be \emph{that bad}. Can the presence of options really be so harmful as to justify their rejection or even annihilation? The death of Buridan's ass, if at all relevant, would stretch the point about the dangers of having options, and how hard it can be to make up one's mind. Everyone agrees that the attractive transformation properties of quantities like $\small{\textsf{C}}$ or $F$ can be interesting, important, physically relevant. Surely that can't be the point; it must be the fate, the contents of the annulus---which already contains what's needed to propagate an influence from the solenoid to the wavefunction. Does it really contain \emph{too much}? Must \emph{too much} really amount to \emph{nothing at all}? Does one really have to dismiss the electromagnetic potential $A$ because many potentials $[A]$ correspond to a given $F$ and $\small{\textsf{C}}$? One can appreciate or even prefer quantities with attractive transformation properties; but is that a reason to \emph{proscribe} equivalence classes altogether? To try to do physics without them? But this concerns (I1) more than (I3)---which has its own embarrassing riches.

I mainly associate the holonomy interpretation (I3) with Richard Healey\footnote{\citet{Healey97, Healey01, Healey04, Healey07}} and Holger Lyre.\footnote{\citet{Lyre, Lyre02, LyreBuch, LyreSHPMP}.  Gordon \citet{Belot1998} attaches ontological importance to holonomies---without, however, \emph{preferring} them to potentials. See also \citet{WuYang}, \citet{Myrvold}.} Their interpretations do differ, and may even evolve; but such ideological `inhomogeneities' complicate the study of almost any doctrine, be it the Copenhagen interpretation or Marxism or Christianity or even spacetime substantivalism. I'll sometimes speak of a generic ``holonomist,'' sometimes more specifically of Healey or Lyre; and must already apologize for any inaccuracies in the attributions. The `philological,' `exegetical' issues---what exactly the literature says---here are arduous and intricate. I cannot claim to have dealt with them all, and leave any outstanding subtleties to more able exegetes. My concerns are chiefly interpretational, philosophical, foundational, in any case not philological; which admittedly raises the secondary issue of exactly how the interpretations considered here relate to those in the literature. They will not, I hope, be altogether unrelated, even if I sometimes idealise and simplify.

Whatever doubts or ambiguities there may be about the holonomy interpretation(s), I think one can at least say this: The holonomist, grudging to pick an individual out of the equivalence class $[A]$,\footnote{\citet{Healey99} p.~444: ``The main problem with ONE TRUE GAUGE is epistemological: the theory itself entails that we could never have any evidence that the TRUE GAUGE was ONE rather than ANOTHER.''} first dismisses the electromagnetic potential as a physically meaningless mathematical fiction;\footnote{See \citet{NTW} on the status of structure often called ``superfluous.''} and then wants to transcend the individuals with a loop integral which, being common to them all, is awarded an appropriate---presumably ontological---primacy.\footnote{\emph{E.\ g}.\ \citet{Healey07} p.~51: ``If the value of the vector potential $A_\mu$ at each space-time point $x$ in a region does not represent some qualitative intrinsic physical properties in the vicinity of $x$, it may be that some function of its integral around each closed curve $C$ in that region does represent such properties of or at (the image of) $C$. [\,\dots] Since the gauge dependence of the vector potential made it hard to accept Feynman's view that it is a real field that acts locally in the Aharonov-Bohm effect, there is reason to hope that a gauge-invariant function of its line integral around closed curves might facilitate a local account of the action of electromagnetism on quantum particles in the Aharonov-Bohm effect and elsewhere.''} The versions of the interpretation may then differ as to exactly what philosophical capital is to be made out of the holonomic entities.

Against the holonomist it can be argued that appropriate electromagnetic properties \emph{can} be assigned to the points of the annulus (\S\ref{connection}); and that an individual loop would have to picked out of its \emph{own} equivalence (homotopy) class, leaving us to wonder why one equivalence class should be any worse than the other (\S\ref{duality}).

A central notion here will be \emph{measurability}: $F$ and $\small{\textsf{C}}$ can be measured, but not $A$---for the time being at any rate. Indeed measurability is a complicated matter: what's unmeasurable today may not be tomorrow (or \emph{vice versa}); it depends on the current state of science, technology, ingenuity, economics and so forth; resources, instruments, capabilities and possibilities available in one spacetime region may not be in another. I'll avoid the most absolute notion of measurability, as being too abstract to accept, and will sometimes include a specification in square brackets: unmeasurable[today], for instance, or measurability[Tuesday] or [with respect to theoretical stipulation $S$], unmeasurability[given the resources available in spacetime region $R$] or [with respect to instrument $\iota$]. Different contexts require different notions of measurability; no notion will be given an absolute primacy, which transcends context. As to $A$, one should really say something like ``$A$ is unmeasurable[now]'' or ``\hspace{-1pt}$A$ is unmeasurable[in the current state of science and technology].'' How do we know that ``\hspace{-1pt}$A$ will never be measured'' or ``\hspace{-1pt}$A$ is unmeasurable \emph{in principle}''? \citet[p.~367]{Maudlin} writes that ``since potentials which differ by a gauge transformation generate identical effects, no amount amount of observation could reveal the ONE TRUE GAUGE,'' which I take to mean ``since potentials which differ by a gauge transformation generate identical effects \emph{for the time being}, no amount amount of observation[now] could reveal the ONE TRUE GAUGE.''\footnote{\emph{Cf}.\ \citet{Healey07} pp.~113-4: ``one cannot rule out a future extension of a Yang-Mills gauge theory that permits observations whose results depend on the existence of a privileged gauge [\,\dots]. If that were to happen, then his observations would discriminate in favor of an interpretation of the gauge theory that commits it to such a privileged gauge, and against a holonomy interpretation. This has not yet happened. But since we cannot be sure that it never will, it seems that we are in no position to answer the question as to whether a holonomy interpretation is correct.'' \citet[footnote 17]{Belot1998} seems to acknowledge the possibility of physically different but empirically indistinguishable potentials $A$, $A'\in[A]$---which makes the empirical indistinguishability appear particularly contingent, perhaps even temporary. See also \citet{AharonovBohm} p.~491: ``we must be able to define the physical difference between two quantum states which differ only by a gauge transformation'' and \citet{Healey09}.\label{four}}

This is related to the matter of \emph{invariance under certain transformations}. The point is not that $A$ can be transformed whereas $F$ and $\small{\textsf{C}}$ cannot (for they can); but that $A$, being unmeasurable[for now], can be subjected to a transformation,
\begin{equation}\label{gauge}
A\mapsto A'=A+d\lambda\textrm{,}
\end{equation}
to which $\small{\textsf{C}}$ and $F$, though functions of $A$, are indifferent. Whereas it makes sense to say that ``$F$ (or $\small{\textsf{C}}$) is gauge invariant,'' the meaning of ``$A$ is not gauge invariant'' or ``$A$ is gauge dependent'' is less clear. Is it meant that $A$ can be transformed? Of course it can---but so can $\small{\textsf{C}}$ and $F$, and in many different ways: $\small{\textsf{C}}\mapsto \small{\textsf{C}}+\zeta$, $F\mapsto 2F$ \emph{etc}. ``$A$ is gauge dependent'' may mean something like ``$\small{\textsf{C}}$ and $F$, which are functions of $A$, are measurable, unlike $A$ itself; and $A$ can be subjected to transformations that leave $\small{\textsf{C}}$ and $F$ unchanged.''\footnote{Lyre has it the other way around---the symmetry \emph{comes first}, $A$ is unmeasurable as a result: (2001) p.~S377 ``\textbf{The Reality of Gauge Potentials.} ``Only gauge-independent quantities are observable.''''; p.~S379 ``holonomies are gauge-independent quantities and therefore appropriate candidates of observable entities''; (2002) p.~82 ``Der Eichsymmetrie zufolge lassen sich Eichpotentiale nicht direkt beobachten -- nur eichinvariante Größen können observabel sein. [\,\dots] In den Eichtheorien sind diejenigen Entitäten, denen aufgrund observabler Konsequenzen Realstatus zugebilligt werden muß, einerseits klarerweise nur bis auf Eichtransformationen festgelegt [\,\dots].''} Shorthand I suppose, but not of the clearest sort.

The transformation (\ref{gauge}) has to be understood in conjunction with the associated phase transformation
\begin{equation}\label{phase}
\psi\mapsto e^{i\lambda}\psi\textrm{,}
\end{equation}
as \citet{Leeds} has rightly emphasized. But it remains true that the gauge dependence of $A$ has above all to be understood in terms of the observability of $F$ and $\small{\textsf{C}}$, and their indifference to (\ref{gauge}). If the $\lambda$ of (\ref{gauge}) \& (\ref{phase}) were fixed by measurement (of phase \emph{or} gauge), both freedoms would disappear together.

\citet{Healey07} devotes much attention to quantized non-Abelian Yang-Mills theory. I only\footnote{Except in \S\ref{connection}, but only for clarity of mathematical exposition.} consider electrodynamics---and hence $\mathbb{U}\textrm{(}1\textrm{)}$, rather than the non-Abelian structure groups of general Yang-Mills theory---without quantization (beyond the introduction of a wavefunction). 

\section{The Aharonov-Bohm effect\label{AharonovBohm}}
A few sentences about the Aharonov-Bohm effect: A wavefunction is split into two, and these, having enclosed a (simply-connected) region $\omega$ containing a solenoid, are made to interfere on a screen. The enclosing wavefunction is sensitive to any enclosed electromagnetism inasmuch as the electromagnetic potential $A$ contributes a phase
$$\exp i\oint_{\partial\omega}\hspace{-3pt}A$$
to (the wavefunction along) the boundary $\partial\omega\equiv\sigma_0$ and hence to the interference pattern on the screen. The electromagnetism on $\omega$ is related to the circulation around the boundary by Stokes's theorem
$$
\small{\textsf{C}}=\oint_{\partial\omega}\hspace{-3pt}A=\iint_{\omega}dA.
$$
The electromagnetic field\footnote{It is perhaps easiest to think of $F$ as a purely \emph{magnetic} field $\mathbf{B}$ produced by the current density $\mathbf{J}=d*\mathbf{B}$ in the solenoid.} $F=dA$ produced by the solenoid is confined\footnote{\citet{Mattingly07} argues that the magnetic field \emph{is} present on the annulus $\alpha$, since its many components are. Indeed they cancel by addition, but one can wonder about the physical meaning of the sum.} to a middle region $\mu\subset\omega$ surrounded by an isolating region\footnote{Viewing $\mu$ and $\omega$ as concentric discs, which is convenient, makes their difference an annulus.} $\alpha=\omega -\mu$ where $F$ vanishes but not $A$. Usually there's just one (cylindrical) shield, around the solenoid; but for the clean delimitation of an (arbitrarily large) intermediate annulus it can be useful to think of \emph{two} coaxial cylindrical shields: one keeping the electromagnetic field \emph{in} inside a larger one keeping the wavefunction \emph{out}.\footnote{In fact there are not two but three `discs' or `circles' or `loops': first, the support of $F$; then, the shield delimiting the annulus on the outside; and finally, the loop running through the wave-function. To simplify, I have conflated the last two. If one would rather distinguish, there is an `integration loop' outside a circle `keeping the wave-function out' (outside the support of $F$).}

Varying the current through the solenoid changes the arbitrarily distant interference pattern, which is perhaps surprising.

\section{The holonomy interpretation(s)\label{three}}

The holonomy interpretation involves appropriate versions of three ideas:
\begin{enumerate}[label=(\roman*)]
\item $[A]=[A+d\lambda]_{\lambda}=d^{-1}F\textrm{ is a class, full of individuals}$\label{idea1}
\item the circulation \small{\textsf{C}} \normalsize is the same for all individuals of the classes $[A]$ and $[\sigma_0]$\label{idea2}
\item transformation properties, symmetries are ontologically relevant.\label{idea3}
\end{enumerate}
But the holonomist seems to---indeed has to---go beyond all three. One can hardly contest \ref{idea1} or \ref{idea2}, which are purely mathematical; and ideas\footnote{See \citet{AfriatCaccese} p.~18.} along the lines of \ref{idea3} go back to \citet{Cassirer} or perhaps \citet{Einstein1916, Quantensatz}\footnote{Einstein sometimes---but not always---gives the impression that transformation properties are ontologically \emph{ir}relevant, \emph{e.g}.\ (1918b) p.\ 167: ``[Levi-Civita] (und mit ihm auch andere Fachgenossen) ist gegen eine Betonung der Gleichung [$\partial_{\nu}(\mathfrak{T}^{\nu}_{\sigma}+\mathfrak{t}^{\nu}_{\sigma})=0$] und gegen die obige Interpretation, weil die $\mathfrak{t}^{\nu}_{\sigma}$ keinen \so{Tensor} bilden. Letzteres ist zuzugeben; aber ich sehe nicht ein, warum nur solchen Größen eine physikalische Bedeutung zugeschrieben werden soll, welche die Transformationseigenschaften von Tensorkomponenten haben.''} or even \citet{Klein}; nor can the point be anything like \emph{the world is made of abstract structures} or \emph{abstract structures should be taken seriously},\footnote{On structural realism see for instance \citet{Poincare02, Poincare05}, \citet{Cassirer}, \citet{Russell}, \citet{Worrall}, \citet{FrenchLadyman}, \citet{French}.} which may be right but not new, either; nor can it be a mere extension of old ideas (about invariance or structures) to yet another theory.\footnote{Can the point just be that \emph{such and such a relationship between invariance and physical reality applies not only to theories} $T_1, T_2,\dots,T_N$, as is well known, \emph{but even to} $T_{N+1}$ (which resembles $T_1,\dots,T_N$ in all relevant ways)?} 
One could argue about \emph{necessity} (``physically real quantities have to be invariant''?) as opposed to \emph{sufficiency}\footnote{\emph{Cf}.\ \citet{Healey97} p.~34: ``since $S\textrm{(}C\textrm{)}$ is gauge-invariant, it may readily be considered a physically real quantity.''} (``\emph{all} invariant quantities are physically real''?), but surely that's a much more abstract issue that does not have to be discussed amid all the intricate peculiarities of (even Abelian) Yang-Mills theory. Necessity is sometimes quite plausible; sufficiency may be a bit strong, but frankly I don't think that's the real issue here. So exactly what is the issue? An appropriate version of \ref{idea3} is combined with \ref{idea1} to rule out\footnote{\citet{Lyre} p.~S377, \citet{LyreSHPMP} p.~665: ``realists can hardly be satisfied by the gauge dependence of entities as imminent in the A-interpretation [\,\dots].'' \citet{Healey97} p.~22: ``there is reason to doubt that the magnetic vector potential is a physically real field, since $\mathbf{A}$ is not gauge-invariant, unlike the magnetic field $\mathbf{B}$ [\,\dots].'' \citet{Healey99} p.~445, \citet{Healey01} pp.~435-6, 454, \citet{Healey07} pp.~25-6, pp.~55-6: ``If there are new localized gauge properties, then neither theory nor experiment gives us a good grasp on them. Theoretically, the best we can do is to represent them either by a mathematical object chosen more or less arbitrarily from a diverse and infinite class of formally similar objects related to one another by gauge transformations, or else by this entire gauge-equivalence class'' and p.~118. Healey's claim that \emph{theory itself} (rather than experimental limitations) rules out a choice of gauge is answered (by himself) in footnote \ref{four} above. \emph{Cf}.\ \citet{Maudlin} pp.~366-7, \citet{Leeds} p.~610, \citet{Mattingly} p.~251, \citet{Healey09} p.~707: ``It is especially hard when it is the theory itself that provides our only initial access to those features of situations it represents by newly introduced structures---hard, but not impossible.''\label{footnote19}} the physical reality of the potential $A$, and an appropriate (perhaps the same) version of \ref{idea3} is combined with \ref{idea2} to favour loops and/or anholonomies, somehow or other. But that's not enough, the holonomy interpretation can hardly stop there, there has to be more. Something along the lines of \emph{a world made of loops}? That would be a bold, original, interesting idea---even more interesting if it were right. Let us try to understand what exactly the holonomist may have in mind.

\subsection{Is an integral a property of its domain of integration?}
The holonomist could argue as follows: To avoid the nonlocality of (I1) there has to be \emph{something} electromagnetic in the annulus $\alpha$. But what electromagnetic entities are available there? Again, the holonomist\footnote{Unlike Healey, who denies any physical reality to the class $[A]$ (and all its individuals), \citet{Lyre02} seems to have no objection to the equivalence class; p.~82: ``Eine genauere Analyse zeigt jedoch, dass nicht den Eichpotentialen, sondern nur Äquivalenzklassen von Potentialen [\,\dots] bzw.\ den so genannten Holonomien ontologische Significanz zukommt [\,\dots].''} dismisses $A$, or even all of $[A]$, by combining \ref{idea1} with an appropriate version of \ref{idea3}. What's left in $\alpha$? The circulation $\small{\textsf{C}}$, the (`once around the solenoid'\footnote{By \emph{hoop} I'll always mean this homotopy class in particular (rather than the classes corresponding to `around twice' or `around thrice' \emph{etc.}).}) hoop and the loops it contains. As \emph{electromagnetic property present in} $\alpha$ the holonomist can pick $\small{\textsf{C}}$, appealing to \ref{idea2} together with an appropriate version of \ref{idea3}.\footnote{\citet{Healey07} p.~105: ``The non-localized gauge potential properties view is motivated by the idea that the structure of gauge potential properties is given by the gauge-invariant content of a gauge theory. The most direct way to implement this idea would be to require that the gauge potential properties are just those that are represented by gauge-invariant magnitudes. [\,\dots] While the vector potential $A_\mu$ is gauge dependent, its line integral $S\textrm{(}C\textrm{)}=\oint_CA_\mu dx^\mu$ around a closed curve $C$ is gauge invariant.} Now that a property has been chosen for $\alpha$, there remains the issue of what it belongs to. Assigning $\small{\textsf{C}}$ to points, or even larger regions, wouldn't make sense, as an entirely gratuitous and uninteresting constant would result. What else could the electromagnetic property $\small{\textsf{C}}$ belong to? A clue is provided by the following trivial example: The circumference $2\pi r$ of a circle is a property of the `domain of integration,' the perimeter. Could the integral $\small{\textsf{C}}$ also belong to its domain of integration,\footnote{The \emph{domain of integration} is often a set. Here a set would work, but something weaker is enough: the notion of `once around the solenoid.'} in much the same way? In our electromagnetic case, what is the domain of integration? A circle, as in the trivial example? Almost---what's needed is a circle up to appropriate (`homotopic') deformations, in other words a hoop. So one holonomy interpretation could amount to this: The integral $\small{\textsf{C}}$ is a property of its domain of integration (just as the circumference was a property of the perimeter).

But can the assignment of an integral to its domain of integration really help us understand the Aharonov-Bohm effect? Can it fill $\alpha$ with a `carrier' medium that deals with the problem of non-locality by transmitting the influence from the solenoid to the wavefunction? One can wonder. So the holonomist may have to go beyond that assignment of an integral to its domain of integration, and make physical (or metaphysical or ontological) claims about the loops in the domain of integration---maybe along the lines of \emph{loops are real} or \emph{loops really exist} or \emph{since loops do the work, they must really be there} or even \emph{the world is made of loops}.

This is a position that, whatever its origin, is at least worth mentioning. I think it resembles Healey's version, or one of Healey's versions, of the holonomy interpretation,\footnote{The following may be relevant: \citet{Healey01} p.~449: ``What is distinctive is not the properties represented by holonomies but the nature of the object whose properties they are. On the present view, holonomies represent global properties of a loop that are not determined by any intrinsic properties of the points on that loop.'' \citet{Healey07} p.~xviii: ``In the simplest case (classical electromagnetism interacting with quantum particles) such an account ascribes properties to (or on) a loop of empty space that are not fixed by properties of anything located at points around the loop [\,\dots].'' P.~31: ``But if the holonomies directly represent electromagnetism and its effects, then there is still a sense in which the action of electromagnetism on the electrons is not completely local, since holonomies attach to extended curves rather than points.'' P.~56: ``only gauge-invariant functions of these mathematically localized fields directly represent new electromagnetic properties; and these are predicated of, or at, arbitrarily small neighborhoods of \emph{loops} in space-time---i.e. oriented images of closed curves on the space-time manifold.'' P.~74: ``This makes it plausible to maintain that what an SU(2) Yang-Mills theory ultimately describes is not a localized field represented by a gauge potential, but a set of intrinsic properties of what I have simply called loops [\,\dots].'' P.~106: ``we arrive at the view that non-localized EM potential properties in a region are represented by the holonomies [\,\dots] of all closed curves in the region [\,\dots]. This is the interpretation of classical electromagnetism I shall defend.'' P.~118: ``One can reformulate the theory as a theory of holonomy properties, so that it does not even appear to mention localized gauge potential properties.'' P.~185: ``gauge potentials directly represent no localized gauge properties, but rather indirectly represent non-localized holonomy properties.'' P.~220: ``[\,\dots] the Aharonov-Bohm effect and other related effects provide vivid examples of physical processes that seem best accounted for in terms of non-localized holonomy properties [\,\dots].'' P.~221: ``Should we believe that non-separable processes involving non-localized holonomy properties are responsible for phenomena like the Aharonov-Bohm effect? This belief may be encouraged by the predictive successes consequent upon introducing classical electromagnetism into the quantum mechanics of particles.'' P.~225: ``This reinforces the conclusion that the evidence for contemporary gauge theories lends credence to the belief that these describe non-separable processes, while nothing in the world corresponds to or is represented by a locally defined gauge potential.'' See also the last paragraph of the book pp.~227-8.\label{HealeyFN}} but I may be wrong. Rather than dwell on the philological issue of exactly how this first holonomy interpretation relates to ideas from the literature, I'll move on to a second holonomy interpretation.

\subsection{Is the world made of loops?\label{loops}}
This second holonomy interpretation is more straightforward than the first. Briefly, it emphasises the ontological primacy of loops (without bothering to point out that an integral belongs to its domain of integration): \emph{Some} physical entity has to do the job, convey the influence; if the embarrassing riches of $[A]$ are dispensed with, what's left in the annulus $\alpha$? Loops---or rather the (more discreetly) embarrassing riches of the homotopy class $[\sigma_0]$. This second interpretation can be attributed to \citet{Lyre},\footnote{P.~S379: ``We [\,\dots] should consider holonomies as physically real.'' P.~S380: ``We may very well represent the physically significant structures in an ontological universe consisting of matter-fields, gauge field strengths, and holonomies.''} \citet{LyreBuch},\footnote{P.~116: ``Die Entitäten der Eichtheorien sind Holonomien [\,\dots].''} \citet{LyreSHPMP},\footnote{P.~665-6: ``we may still consider holonomies as object-like entities, but to such an extent that our notion of an object becomes highly abstract.'' Lyre (p.~658) sees holonomies as ``basic entities'' and ``genuine entities.'' \citet{Belot2003} p.~216 has a similar position---without, however, going so far as to claim that loops are \emph{more} real: ``holonomies [\,\dots] are well-defined quantities on the spaces of states of the standard formulations of Yang-Mills theories. If it is accepted that these theories describe reality, does not it follow that the quantities in question are as real as any others?'' See also \citet{Belot1998} p.~544: ``we must also consider closed curves in space to be carriers of the electromagnetic predicates'' and the final paragraph pp.~553-4.} or again to \citet{Healey01},\footnote{P.~448: ``it is the holonomies that represent the real physical structures in a gauge situation, rather than any particular bundle connection that is compatible with them.''} \citet{Healey07}.\footnote{P.~30: ``Suppose instead that one takes the holonomies themselves directly to represent electromagnetism and its effects on quantum particles.''}

In \S\ref{duality} I'll argue as follows: $A$ and $\sigma_0$ are related by a significant duality, so that however one feels about the class $[A]$ (or its elements), it is no worse than the hoop $[\sigma_0]$ (or its elements). I'll think of $A$ as a set $\{\sigma_1,\dots,\sigma_N\}$ of level curves, which can indeed be deformed (by (\ref{gauge}))---but so can $\sigma_0$. The deformability of $\sigma_1,\dots,\sigma_N$, or rather $\sigma_0,\dots,\sigma_N$, is neither here nor there, and shouldn't be used to rule out the reality of $A$ in particular.

I also explore (\S\S \ref{garden},\ref{electrostatics}) the relationship between the measurability of a quantity (say $A$) and the ontic status of a boundary; can the measurement of one entity undermine the physical reality of another, ferrying it off to a shadier realm of mathematical fictions?

\subsection{Summary}
Before turning to my objections, I can try to summarize the holonomy intepretation(s): The electromagnetic field $F$ and inverse image $[A]=d^{-1}F$ are measurable, but not the individual potential $A$---whose physical reality is therefore questionable. But surely the Aharonov-Bohm effect has to be conveyed by \emph{something}. If the potential isn't really there, what's left? The solenoid, and the electromagnetic field it contains, are (arbitrarily) far from the wavefunction and the screen on which the effect is seen. The circulation $\small{\textsf{C}}$, which determines the interference pattern, has a promising indifference to (\ref{gauge}); but $\small{\textsf{C}}$ is just a number, not enough on its own to convey or account for the effect---something more is presumably sought. The number is obtained by integrating any $A\in[A]$ around any $\sigma_0\in[\sigma_0]$; having ruled out $A$,  the holonomist attributes reality to $\sigma_0$ (or $[\sigma_0]$) instead.

As a compressed statement we can adopt $\mathfrak{S}$: \emph{Since $[A]$ is a class it cannot be taken seriously; physical meaning should be attributed to the class $[\sigma_0]$ instead}.

\section{Objections}
\subsection{The connection\label{connection}}
Again, if it were possible to fill the annulus $\alpha$ uncontroversially with appropriate electromagnetic properties, there would be no reason to look beyond the potential interpretation (I2). Indeed the holonomist's first step, to make the abstractions of (I3) seem indispensable, is to dismiss $A$ as a meaningless mathematical fiction: $A$ does not assign a property $A_x$ to the physical point $x$ because there are so many other potentials in $[A]$. The move can be countered by producing an electromagnetic function $\varphi$, closely related to $A$, that does assign a property $\varphi_x$ to every $x$.

Structure often called `surplus'---gauge, components \emph{etc}.---can sometimes be transcended by appropriate geometrical abstractions; so one can take the one-form $A=A_\mu dx^\mu$ rather than its components $A_\mu$ with respect to this or that basis $dx^\mu$, the tensor
$$T=T^\mu_{\nu\sigma}\partial_\mu\otimes dx^\nu\otimes dx^\sigma$$
rather than its components $T^\mu_{\nu\sigma}$ and so on. Can something similar be attempted here? \citet[p.~445]{Healey99} writes that ``one can represent the gauge potential in a gauge-independent way as a connection one-form on a principal fiber bundle [\,\dots].'' Since the fiber of the principal bundle\footnote{We can take the principal bundle to be trivial, in other words a Cartesian product, with no relevant loss of generality. \emph{Cf}.\ \citet{Healey99} p.~445: ``a gauge potential is more appropriately represented mathematically not as a vector field defined on spacetime points, but rather as a connection one-form on a principal fiber bundle over (some region of) spacetime. It is only when this bundle representation is trivial that the older vector representation of the potential is even possible. That is always the case for Maxwellian electromagnetism defined over Minkowski spacetime, but it is not true if one allows for the possibility of magnetic monopoles.'' The base space $M$ is flat spacetime (or even just space; time has an uninteresting rôle here). } $P=M\times G$ is the structure group $G$, which represents local gauge transformations and freedom, the way may not be as ``gauge-independent'' as he suggests, but we can try to get around the problem. By ``gauge-independent'' Healey no doubt means something like `invariant' or `geometrical' or `equivariant': the dependence on $G$ (which generalises the $\mathbb{U}\textrm{(}1\textrm{)}$ phase transformation (\ref{phase})) can be balanced by something along the lines of (\ref{gauge}) to produce a higher invariance expressed in the Lagrangian or field equations or elsewhere. But even if the local dependence on the structure group can thus be overcome somewhere or other by means of appropriate compensation, $G$ remains undeniably present on the principal bundle. So we have to find a way of descending from the principal bundle to $M$, to whose points we want to assign the electromagnetic property. There will be no attempt to do away with gauge freedom altogether; rather, it will be shown that such freedom by no means prevents the assignment of an appropriate electromagnetic propert\emph{y} to each point of the annulus.

Properties in modern physics, when unpacked, can have plenty of internal structure: many apparently simple properties are in fact functions, equivalence classes, tensors \emph{etc}. Here, nothing prevents us from reorganising the elements of $P$ so as to turn the fiber $G$ into the domain of a function $\varphi$ on $M$, whose value
$$
\varphi_x:G_x\rightarrow [\mathbb{H}_xP];\hspace{3pt}g\mapsto\varphi_x\textrm{(}g\textrm{)}=\mathbb{H}_{\textrm{(}x,g\textrm{)}}P
$$
at $x\in M$ would then be the required electromagnetic property, the range
$$
[\mathbb{H}_xP]=[\mathbb{H}_{(x,G_x)} P]=[\mathbb{H}_{\textrm{(}x,g\textrm{)}} P]_g
$$
being the class of horizontal subspaces\footnote{Such a subspace indicates how the element $g\in G_x$ gets \emph{radiated} linearly from $x$ to the surrounding points of $M$; since each element $g$ of each copy $G_x$ of $G$ has to be radiated separately, the connection assigns a horizontal subspace to each point of the principal bundle.} assigned by the connection to $x$, one for every $g\in G_x$. Again, gauge freedom is still there, but at least we have a single electromagnetic property at $x$---\emph{single} by the standards of internal structure typical to physics, to modern physics at any rate. \emph{Cardinality} (or perhaps the grammatical category of \emph{number}---singular, plural) is not invariant under mathematical reformulation: a rearrangement of the principal bundle has taken us from the embarrassing riches of $[A_x]$ to the single propert\emph{y} $\varphi_x$.

Again, the holonomist argument amounts to this: \emph{the individuals are a nuisance and have to be transcended; at a point we're stuck with them, they simply cannot be dispensed with: hence integration, loops} etc. So the crucial move to loops and integration is only justified by the supposed impossibility of getting rid of the individuals of $[A]$ at a point $x$; if they could already be punctually transcended, why bother with integration? Healey \emph{has to} claim that it is \emph{mathematically impossible} to define an object (potential or connection or covariant derivative or whatever) at point $x$ that corresponds---without integration---to the whole class $[A_x]$. His argument\footnote{1999 p.~446, 2001 p.~436ff., 2004 p.~628} seems to be that \emph{since one can always introduce the individuals, since one can always talk about them, they can't be eliminated}. It is undeniably true that the individuals can always be introduced or referred to; indeed one has to be able to introduce them; vertical automorphisms or equivalent operations are always \emph{available}, however abstract or geometrical the formulation. But the individuals can nonetheless be transcended, packed away out of sight, by means of appropriate constructions or reformulations.

\emph{Picking} individuals (gauge fixing) and \emph{transcending} them may look like incompatible solutions. Perhaps; but at worst a choice would be forced---pending which, both remain available.

\subsection{Duality between loops and potentials\label{duality}}
There is a significant duality between loops and potentials; just as a vector $\dot{\sigma}'_0\textrm{(}x\textrm{)}\in T_xM$ and a covector $A\textrm{(}x\textrm{)}$ from the dual space $T^*_xM$ give a number $\langle A\textrm{(}x\textrm{)},\dot{\sigma}'_0\textrm{(}x\textrm{)}\rangle$, the loop\footnote{The unparametrized curve $\sigma_0\subset M$ is the image of the parametrized curve $\sigma'_0:I\rightarrow M\,;\,t\mapsto\sigma'_0\textrm{(}t\textrm{)}$, without its parameter $t$, which is not part of the boundary $\partial\omega$ (where $I\subset\mathbb{R}$ is an appropriate interval and $M$ flat spacetime).} $\sigma_0\equiv\partial\omega$ and potential give a number $\textrm{(}A,\sigma_0\textrm{)}=\small{\textsf{C}}$. Both $A$ and $\sigma_0$ can be deformed without affecting the circulation: the potential according to (\ref{gauge}); a loop can be deformed into any other loop going around the solenoid once. Both could be replaced by their equivalence classes $[A]$ and $[\sigma_0]$, one could even write $([A],[\sigma_0])=\small{\textsf{C}}$.

It will be useful to understand the transformation (\ref{gauge}) more geometrically, as a deformation of the level sets of $A$'s local primitive\footnote{For wherever $A$ is closed it can be written locally as the gradient $A=d\sigma$ of a zero-form $\sigma$.} $\sigma$.\footnote{A similar construction is used in \citet{THS}.} One can first imagine a purely `angular' or `radial' $\sigma$ (with values running from zero to $2\pi k=\small{\textsf{C}}$),\footnote{Such a $\sigma$ cannot be continuous everywhere; we can imagine a single discontinuity, say on the ray with values $\sigma=2\pi nk$, where the integer $n$ is zero then one, $k=\scriptsize{\textsf{C}}/2\pi$ being a constant.} whose level lines are straight rays radiating through the annulus $\alpha$ from the inner disc $\mu$ to the edge $\partial\omega$. A gauge transformation (\ref{gauge}) would then deform the level rays, bending them without making them cross. It is easier to picture the denumerable set $\{\sigma_1, \dots, \sigma_N\}$ of level curves at intervals of $\small{\textsf{C}}/N$ than all of them; they will each be cut once\footnote{Or rather \emph{an odd number of times}, as Jean-Philippe Nicolas has pointed out to me. Crossings in opposite directions cancel, and add nothing to the integral.} by any loop $\sigma_0$ going around the solenoid once.

In this construction we have $N+1$ deformable curves $\sigma_0, \dots, \sigma_N$, which all seem pretty much on the same footing; $\mathfrak{S}$ amounts to the surprising claim that $\mathfrak{S}'$: \emph{Only $\sigma_0$ is physically real (along with its deformations) because the other curves $\sigma_1, \dots, \sigma_N$ can be deformed}. Why should one curve $\sigma_k$ be any better than the others? How about $\sigma_7$? It remains true that $\sigma_0,\dots,\sigma_6,\sigma_8,\dots,\sigma_N$ can be deformed.

To emphasize that loops are no better than $A$, we can even arrange for a gauge transformation to \emph{induce} a loop deformation (thus strengthening the duality): Level rays of unit length determine a unit circle, which will then be `deflated' into a smaller loop by a gauge transformation (\ref{gauge}); to every such transformation there corresponds a different loop $\sigma_0^\lambda$. If a potential subject to (\ref{gauge}) is too flimsy to exist, why should loops also subject to (\ref{gauge}) be any better? Are vectors any more real than the covectors dual to them? Is a Lagrangian ontologically inferior to the Hamiltonian dual to it? Does momentum exist more than velocity?

Healey's agenda is undermined by the possibility of constructing, without integration, an appropriate electrogeometrical object $\varphi$ with a single value $\varphi_x$ at every point $x$. But suppose, for the sake of argument, that such an object \emph{cannot be constructed}, as indeed Healey has to claim. He still has to establish that the individuals of the equivalence class $[A]$ are more troublesome than those of the hoop $[\sigma_0]$. Again, I've argued that there's a significant duality between the two, and that $[A]$ is no worse than $[\sigma_0]$. Or even \emph{somewhat better}: Since the hoop $[\sigma_0]$ is more homogeneous than $[A]$, it makes more sense to pick a potential out of $[A]$ than to pick a loop out of $[\sigma_0]$; the elements of $[\sigma_0]$ all look about the same, whereas $[A]$ contains distinguished elements that stand out from the rest. \emph{Gauge fixing} is an accepted and legitimate practice---but who would ever indulge in \emph{loop fixing}, picking an individual loop out of $[\sigma_0]$, preferring it to the others? (Would \emph{ellipses} be better than other loops? How about roughly octogonal loops?)

\subsection{Vertical drop\label{garden}}
Suppose I can only measure the curl $F=dA$ of my garden's gradient\footnote{Needless to say, most gardens have an exact gradient.} (or infinitesimal vertical drop) $A$ but not the gradient itself---some instruments and experimental possibilities are available, others aren't.\footnote{Again, there can be other instruments and experimental possibilities in other spacetime regions; experimental possibility may or may not be shared between regions.} The indifference of $F=dA+d^2\lambda$ to the exact one-form $d\lambda$ can accordingly be called a \emph{gauge freedom}. The vertical drop or rather circulation
$$
\small{\textsf{C}}=\oint_{\partial\omega}\hspace{-3pt}A=\iint_{\omega}dA
$$
is also indifferent to (\ref{gauge}), where $\partial\omega$ is the boundary of a region $\omega$.

The unmeasurability[in an appropriate spacetime region] of $A$ makes my garden an awkward tangle of real physical loops by producing the gauge freedom which in turn invests the boundary $\partial\omega$ (or $[\partial\omega]$) with physical reality. But to dispatch the loops to the shady regions (populated by innocuous mathematical ghosts) where they obstruct neither gardening nor evening strolls it may be enough to \emph{work out how to measure} $A$; for that could change the status of (\ref{gauge}) and hence of the loops.

We know that peeking can \emph{kill a cat} \citep{Schroedinger}; but here, peeking at a gradient may well \emph{annihilate a loop}!

\subsection{Electrostatics, gravity\label{electrostatics}}
The solenoid, or perhaps the field $F=dA$ it produces, is a source whose `radiation' $A$ is caught by the boundary $\partial\omega$, giving the circulation $\small{\textsf{C}}$.

In electrostatics the source is a charge density three-form $\rho=d\mathbf{E}$, which radiates the electric two-form $\mathbf{E}$ caught by the boundary $\partial\Omega$ of a region $\Omega$ containing $\rho$. Stokes's theorem again holds, and allows us to write
$$\small{\textsf{F}}=\iint_{\partial\Omega}\mathbf{E}=\iiint_{\Omega}d\mathbf{E}=\iiint_{\Omega}\rho.$$
The difference here is that, unlike $F$'s primitive $A$, $\rho$'s primitive $\mathbf{E}$ \emph{is} measurable (and fixed by the condition $\mathbf{E}=*d\varphi$, the electrostatic potential $\varphi$ being a zero-form). $\small{\textsf{F}}$ and $\rho$ can nonetheless be called `gauge invariant,' in the sense that they're indifferent to the gauge transformation
\begin{equation}\label{gauge2}
\mathbf{E}\mapsto\mathbf{E}'=\mathbf{E}+d\beta\textrm{,}
\end{equation}
where $d\beta$ is the curl of a one-form $\beta$. To make the gauge transformation (\ref{gauge2}) physically meaningful, one would have to \emph{forget how to measure} $\mathbf{E}$.\footnote{It is by no means impossible to make a quantity unmeasurable; one can destroy instruments, proscribe know-how, banish specialists and so on. Measurability is as reversible as progress.} Once $\mathbf{E}$ is empirically inaccessible, (\ref{gauge2}) will acquire a different ontic status, and so will the boundary $\partial\Omega$. If we then feel trapped inside an infinite gauge-invariant class $[\partial\Omega]$ of real physical membranes, it may be enough, to dissolve them all, to \emph{remember how to measure} $\mathbf{E}$.

The same applies, \emph{mutatis mutandis}, to Newton-Poisson gravity, where $\rho$ is the mass density, $\varphi$ the gravitational potential \emph{etc}.\footnote{See \citet{THS}.}

The same also applies to the Aharonov-Bohm effect itself: Suppose an ingenious experimenter works out how to measure $A$. That would change the status of (\ref{gauge})---it would be taken less seriously---and hence of $\partial\omega$, which would undergo an ontic transition. We have something like a `law of ontological conservation': if the reality isn't \emph{here} ($A$), and it has to be somewhere, it must be \emph{there} ($\partial\omega$); but if it \emph{is} here, it no longer has to be there \dots

\section{Final remarks}
We all agree that (\ref{freedom}) is a class, full of individuals; and that the integral $\small{\textsf{C}}$ of the electromagnetic potential $A$ around a loop $\sigma_0$ is common to the whole class $[A]$, and to the hoop $[\sigma_0]$. Surely the holonomist has to go beyond a re-statement of those incontrovertible mathematical facts, to make a physical or metaphysical point of some sort. The language used suggests it is an ontological point, involving integrals around loops. The integral here is just a number---about which it seems hard to make interesting ontological claims. Who would dispute that $\small{\textsf{C}}$ is \emph{important} or \emph{physically relevant}? Is the claim that the number \emph{exists}? Or that it represents a physical quantity? If the holonomy interpretation is to go beyond the mere number $\small{\textsf{C}}$, it seems obliged to make ontological claims about \emph{loops}, along the lines of \emph{loops are physically real} or \emph{loops really exist} or \emph{the world is made of loops}.

Picking a potential $A$ out of $[A]$ seems no worse than picking a loop $\sigma_0$ out of the hoop $[\sigma_0]$. There may be reasons to prefer $[A]$ to any particular potential $A$, and to prefer $[\sigma_0]$ to any particular loop $\sigma_0$; but those reasons are not good reasons to prefer $[\sigma_0]$ to $[A]$, or any particular $\sigma_0$ to any particular $A$; nor are they good reasons to prefer $[\sigma_0]$ to any particular $A$ (or $[A]$, for that matter, to any particular $\sigma_0$).

We owe the embarrassing riches of (\ref{freedom}) to the unmeasurability of potentials, about which I have raised questions in \S\S \ref{garden}-\ref{electrostatics}. It is remarkable that the empirical inaccessibility of a differential form $\zeta$ on $\Omega$ should confer physical reality on the boundary $\partial\Omega$. If the `potential' represented by $\zeta$ sooner or later becomes measurable, will the boundary $\partial\Omega$ fade and dissolve?

\vspace{10pt}
\noindent I thank Stefano Bordoni, Ermenegildo Caccese, Claudio Calosi, Adam Caulton, Dennis Dieks, Johan Huisman, Antonio Masiello, Jean-Philippe Nicolas, John Norton and Brian Pitts for valuable comments; and Dimitri Kasprzyk for telling me about \emph{énantiosèmes.}

\end{document}